\documentclass[prl,aps,twocolumn,superscriptaddress,showpacs]{revtex4}

\usepackage{graphicx}

\begin{document}

\title{Incommensurate short-range order in the $S$=1 triangular antiferromagnet NiGa$_{2}$S$_{4}$}

\author{C. Stock}
\affiliation{ISIS Facility, Rutherford Appleton Labs, Chilton, Didcot, OX11 0QX, UK}
\author{S. Jonas}
\affiliation{Department of Physics and Astronomy, Johns Hopkins University, Baltimore, Maryland USA 21218}
\author{C. Broholm}
\affiliation{Department of Physics and Astronomy, Johns Hopkins University, Baltimore, Maryland USA 21218}
\affiliation{NIST Center for Neutron Research, Gaithersburg, Maryland USA 20899}
\author{S. Nakatsuji}
\affiliation{Institute for Solid State Physics, University of Tokyo, Kashiwa, Chiba 277-8581, Japan}
\author{Y. Nambu}
\affiliation{Institute for Solid State Physics, University of Tokyo, Kashiwa, Chiba 277-8581, Japan}
\author{K. Onuma}
\affiliation{Institute for Solid State Physics, University of Tokyo, Kashiwa, Chiba 277-8581, Japan}
\author{Y. Maeno}
\affiliation{Department of Physics, Kyoto University, Kyoto 606-8502, Japan}
\author{J.-H. Chung}
\affiliation{NIST Center for Neutron Research, Gaithersburg, Maryland USA 20899}

\date{\today}

\begin{abstract}

Neutron scattering is used to investigate spin correlations in ultra pure single crystals of the S=1 triangular lattice NiGa$_{2}$S$_{4}$.  Despite a Curie-Weiss temperature of $\Theta_{CW}$=-80(2) K, static ($\tau$ $>$ 1 ns) short range ($\xi_{ab}$=26(3) \AA) incommensurate order prevails for T$>$1.5 K.  The incommensurate modulation ${\bf{Q_{0}}}$=(0.155(3),0.155(3),0), $\Theta_{CW}$, and the spin wave velocity ($c$=4400 m/s) can be accounted for by antiferromagnetic third-nearest neighbor interactions $J_{3}$=2.8(6) meV and ferromagnetic nearest neighbor coupling $J_{1}$=-0.35(9) $J_{3}$.  Inter-plane correlations are limited to nearest neighbors and weakened by an in-plane field.  These observations show that the short range ordered glassy phase that has been observed in a number of highly degenerate systems, can persist near the clean limit.

\end{abstract}


\maketitle

	In geometrically frustrated magnets no spin state optimizes all interactions and consequently spins can remain disordered at temperatures ($T$) well below the Curie-Weiss temperature.~\cite{Ramirez01:13}  The frustration results in high degeneracy as in strongly correlated electronic systems but in a context that is more amenable to analysis and comparison between theory and experiments.  Frustrated magnets thus provides a unique opportunity to explore the new states of matter that can result from high degeneracy and suppression of conventional mean field behaviors.   The simplest frustrated magnet consists of antiferromagnetically interacting spins on a triangular lattice.  For S=1/2 quantum fluctuations play an important role with long-range order only predicted to exist at T=0.~\cite{Wannier50:79,Eggarter75:12,Morita02:71}  In addition for weakly distorted S=1/2 systems close to a metal insulator transition such as $\kappa$-(BEDT-TTF)$_{2}$Cu$_{2}$(CN)$_{3}$, there is evidence of a spin liquid phase with no sublattice magnetization.~\cite{Shimizu03:91} For the classical ($S \to \infty$) limit long range spin order at finite temperatures has been observed in materials such as RbFe(MoO$_{4}$)$_{2}$ and NaMnO$_{2}$.~\cite{Kenzelmann07:98,Giot07:94}  The situation for S=1 is less clear with a resonating valence bond state and a spin nematic phase having been proposed.~\cite{Tsun05:xx,Bhatt06:74,Stoudenmire09:79} Interestingly, there are no experimental examples of isotropic triangular lattice antiferromagnets with S $\leq$ 1 which possess conventional long-range order.  It is therefore important to investigate the magnetic properties in isotropic, stoichiometric S=1 triangular lattice systems, where collective phases may be possible.
	
	NiGa$_{2}$S$_{4}$ consists of two-dimensional (2D) planes of $S$=1 Ni$^{2+}$ magnetic ions occupying the corners of edge sharing equilaterial triangles with Van der Waals forces alone stabilizing the lamellar structure.~\cite{Nakatsuji05:309,Nakatsuji07:99}  While magnetization measurements yield $\Theta_{CW}$=-80(2) K, short-range (SR) order prevails for T$>$1.5 K.   Here we study the low-T magnetic properties of NiGa$_{2}$S$_{4}$ using neutron scattering from high quality single crystals and provide clear evidence for quasi 2D incommensurate SR order.  While proposals for a spin nematic phase may yet be relevant, we show that important aspects of the low energy physics are well described by a $J_{1}$-$J_{3}$ model with strong third nearest neighbor (NN) antiferromagnetic exchange and a weakly ferromagnetic NN exchange.  In this model, the lack of long range would be associated with anomalous quasi-2D domain dynamics.  

	The experiments used the SPINS and BT2 triple-axis spectrometers and the High Flux Backscattering (HFBS) instrument at the NIST Center for Neutron Research.  Sintered pellets and two different single crystal configurations were used.  We used 7 co-aligned single crystals (with a mass of 300 mg) for the (HHL) reciprocal lattice plane and 19 single crystals (1 g) for the (HK0) zone.  NiGa$_{2}$S$_{4}$ crystals were grown by the chemical vapor transport method followed by annealing in a sulphur atmosphere.  ICP analysis showed the stoichiometry to be within 1\% of the nominal composition.  The low-T lattice constants were $a$=$b$=3.624 \AA\, $c$=11.999 \AA\, $\alpha$=$\beta$=90$^{\circ}$ and $\gamma$=120$^{\circ}$.~\cite{Lutz86:533}  

\begin{figure}[t]
\includegraphics[width=85mm]{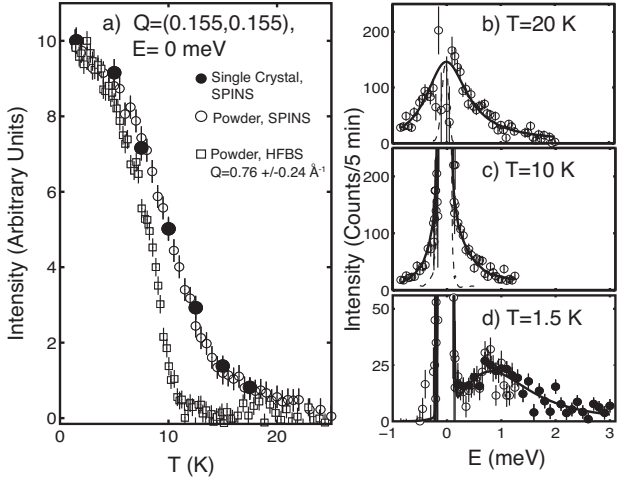}
\caption{$a)$ The $T$ dependence of magnetic neutron scattering at ${\bf{Q}}$=(0.155,0.155,1) for NiGa$_{2}$S$_{4}$ single crystals and powder taken on SPINS with E$_{f}$=3.7 meV.  $b)$-$d)$ The magnetic spectrum at 20 K, 10 K, and 1.4 K. The open and filled circles were taken with the crystals aligned in the (HHL) and (HK0) scattering planes and with the analyzer configured for 11$^{\circ}$ and 5$^{\circ}$ acceptance respectively. The curves defined by the dashed line represents the energy resolution.} \label{temp}
\end{figure}

\begin{figure}[t]
\includegraphics[width=75mm]{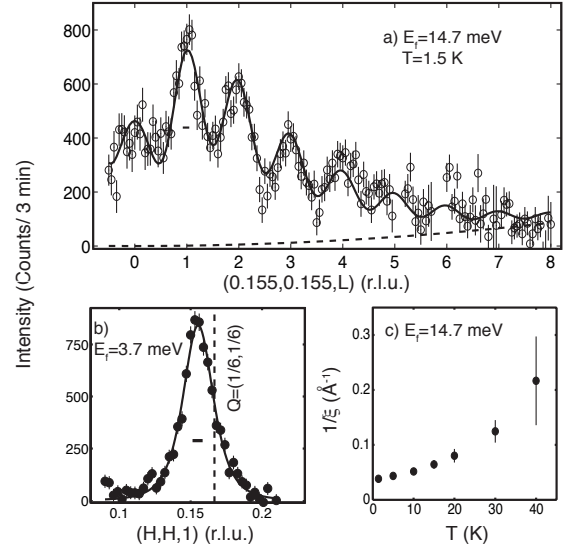}
\caption{$a)$ - the elastic magnetic scattering versus momentum transfer along the L direction using 30 K as a background.  $b)$ - the momentum dependence along the [110] direction with the vertical dashed line marking the commensurate (1/6,1/6) position.  $c)$ - the correlation length versus $T$ measured in 2-axis mode.  The horizontal bars are the resolution full-width.} \label{momentum}
\end{figure}

	We first examine the development upon cooling of static SR spin correlations.   Fig. \ref{temp} $a)$ shows the $T$ dependence of elastic scattering from powder and single crystals acquired with differing energy resolution.  The single crystal and powder data on SPINS show a gradual onset of elastic scattering with decreasing T.  For the powder sample, the increase in the intensity for 0.52$<$Q$<$1.0 \AA$^{-1}$ beyond that observed at T=40 K is a measure of the magnetic elastic scattering.  The single crystal data were obtained through (HH1) scans fitted to a Lorentzian squared to obtain the integrated intensity (Fig. \ref{momentum} $b)$).  On tightening the energy resolution from $\delta E$=80 $\mu eV$ (on SPINS) to 0.8 $\mu eV$ (on HFBS), the characteristic $T$ has decreased substantially and the onset of magnetic intensity upon cooling is sharper.  The backscattering data indicate spin freezing on a timescale which exceeds $\tau$ $\sim$ $\hbar / \delta E$ = 1 ns.  The strong correlation of the onset $T$ with the energy resolution indicates that the elastic scattering in NiGa$_{2}$S$_{4}$ is associated with rapid evolution with $T$ of the time scale for incommensurate spin fluctuations rather than a conventional second order phase transition.  These observations are reminiscent of observations at the spin glass transition in Cu-Mn.~\cite{Murani89:41} The onset $T$ for elastic magnetic scattering for $\delta E =$0.8 $\mu$eV coincides with the broad maximum in magnetic specific heat (C$_{m}$) but is slightly above the 8.5 K susceptibility anomaly.  Similar observations were reported in strongly disordered 2D SCGO.~\cite{Broholm90:65}  For crystalline NiGa$_{2}$S$_{4}$ however, doping at the 1 \% level was found to drastically suppress the low T maximum in C$_{m}$ indicating an impurity concentration well below 1 \% for our crystals, which do display the  C$_{m}$ anomaly.~\cite{Nambu06:75}  These results illustrate that our samples are very pure with minimal disorder present.

	We now discuss inelastic neutron scattering data that probe spin fluctuations versus T.  The magnetic spectrum at the critical wavevector ${\bf{Q_{0}}}=(0.155,0.155,0)$ is displayed in Fig. \ref{temp} panels $b)$ through $d)$.  The data were obtained on SPINS and corrected for a background measured at $\bf{Q}$=(0.2,0.2,1) (open circles) and (0.25,0.5,0) (filled circles). At 20 K, there is \textit{no truely elastic scattering} and the inelastic spectrum shows a broad peak in energy that can be described by a spin relaxational lineshape $S({\bf{Q}},\omega) \propto [1/(1-e^{-\hbar\omega/k_{B}T})]\omega/(\Gamma^{2}+\omega^{2})$ with $\hbar \Gamma$=0.4(1) meV.  At lower T, strong elastic (on the timescale of our measurement) magnetic scattering develops with two timescales; a resolution limited central peak and a broad relaxational response perhaps associated with exchange or single ion anisotropy. The inelastic spectra indicate slowing spin fluctuations upon cooling, culminating in the development of a disordered staggered magnetization frozen on a time scale that exceeds $\sim$ 1 ns.     

	The momentum dependence of the elastic scattering is summarized in Fig. \ref{momentum}.  Panels $a)$ and $b)$ illustrate correlations along the $c$-axis and within the $a-b$ plane respectively.  The solid line in panels $a)$ and $b)$ are fits to the following formula, which represents SR correlations within the $a-b$ plane and correlations between NN triangular lattice planes.   

\begin{eqnarray}
I({\bf{Q}}) =  C {{(\gamma r_{0})^{2}}\over{4}} {{{m}^{2}}}  (1+2\alpha\cos({\bf{Q}}\cdot{\bf{c}})) 
\label{formula1}
e^{-\langle u \rangle ^{2}Q^{2}} \\
g^{2} f^{2}(Q) {{(1-\sin^{2}\theta\cos^{2}\phi\hat{q}_{\perp}^{2}-\cos^{2}\theta\hat{q}_{\parallel}^{2})} A^{*} \xi^{2}/\pi \over {(1+(\xi|{\bf{Q}}_{ab}-{\bf{Q}}_{0}|)^{2}})^{2}} \nonumber
\end{eqnarray}

\begin{figure}[t]
\includegraphics[width=68mm]{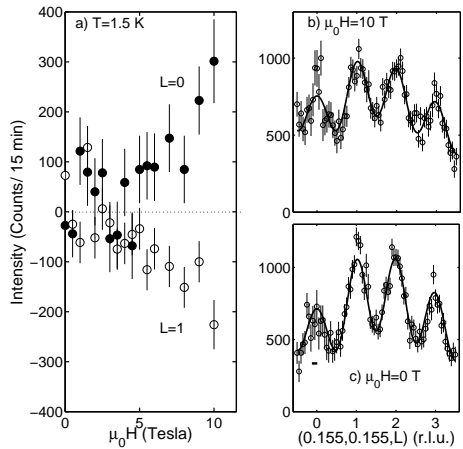}
\caption{Magnetic field dependence of the neutron scattering at ${\bf{Q}}$=(0.155,0.155,L) measured on SPINS.  $a)$ - the field dependence of the change in peak intensity at L=0 and L=1. $b)$ and $c)$ - the intensity versus L at 10 and 0 $T$ with 25 K as a background.  The horizontal bars are the resolution full-width.} \label{field}
\end{figure}

\noindent $f^{2}(Q)$ is the Ni$^{2+}$ form factor, $\xi$= 26(3) \AA\ is the correlation length within the $ab$ plane, $\langle u^{2} \rangle$ is the mean squared displacement of Ni (taken to be 0.005 \AA$^{2}$)~\cite{Jones:unpub}, ${\bf{Q}}_{ab}$=(H,K) is the in-plane wavevector, $g$ is the Lande factor, $(\gamma r_{0})^{2}$ is 0.292 barn, $A^{*}$ is the area of the Brillouin zone, and ${\bf{m}}=m(\sin\theta\cos\phi\hat{x}+e^{i\psi_{1}}\sin\theta\sin\phi\hat{y}+e^{i\psi_{2}}\cos\theta\hat{z}$) with $\hat{x}$ $\parallel$ to ${\bf{Q_{0}}}$=(0.155,0.155,0) and $\hat{z}$ $\parallel$ $c$.  In terms of ${\bf{m}}$, the time averaged spin on site ${\bf{r}}$ is given by ${\langle\bf{S}}({\bf{r})}\rangle=2 Re\{ {\bf{m}} e^{i\bf Q\cdot r}\}$. While ${\bf{S}}({\bf{r}})$ depends on the phase $\psi_{1}$ the magnetic scattering intensity for an unpolarized beam (Eqn. \ref{formula1}) does not.  We assume the overall spin configuration (after domain averaging) is invariant under symmetry operations of the paramagnetic phase, eliminating effects of $\psi_{2}$.  The parameter $\alpha$=0.16(3) indicates nearest-neighbor inter-plane ferromagnetic correlations.  The calibration constant $C$ was determined from 12 nuclear Bragg peaks ranging from (001) to (114) and indicates $\langle S \rangle^{2}$=0.26(5).  The errorbar on $\langle S \rangle^{2}$ is due to uncertainty of the crystal mosaic. The strongly reduced sublattice magnetization ($S$=1) and is less than reported for a powder sample~\cite{Nakatsuji05:309} and indicates strong quantum fluctuations.~\cite{Igarashi92:46}  The dashed line in panel $a)$ is a background term $\propto$ $Q^{2}$ resulting from non-cancellation of the Debye Waller factor affected elastic incoherent nuclear scattering. A Lorentzian squared was chosen to describe the momentum dependence of the in-plane correlations as it is normalizable in 2D and has been used to describe SR correlations in random field magnets.~\cite{Birgeneau83:28}  

	Eqn. \ref{formula1} convolved with the resolution fits the complete $L$ dependence.  The broad maxima for integer $L$ indicate ferromagnetic correlations between NN planes. Two spin configurations fit the single crystal data equally well.  An out of plane spiral with $\phi$ $\equiv$ $0^{\circ}$ and $\theta = 57(5) ^{\circ}$ and an in-plane spiral with $\phi =41(5)^{\circ}$ and $\theta$ $\equiv$ $90^{\circ}$.  Higher angle peaks in the powder data are better described with spins in the basal plane.~\cite{Nakatsuji05:309} 

	We now examine in-plane spin correlations.  Fig. \ref{momentum} $b)$ shows a scan along the (HH1) direction.  Instead of the conventional 120$^{\circ}$ structure, the data are more closely described by a 60$^{\circ}$ structure as expected for dominant third NN interactions ($J_{3}$) consistent with first principles calculations.~\cite{Mazin07:76}  In addition, the peak is clearly shifted below the ${\bf{Q}}$=(1/6,1/6) commensurate position.  As detailed below, the incommensurability may result from competing first and third NN interactions.

	The $T$ dependence of the inverse in-plane $\xi$ is plotted in Fig. \ref{momentum} $c)$ as obtained on the BT2 spectrometer in two-axis mode with E$_{i}$=14.7 meV.  The data represent an approximate measurement of the instantaneous correlation length.  $\xi$ increases upon cooling saturating at 26(3) \AA\ with no finite-T anomaly as would be expected at a conventional second order phase transition. 


\begin{figure}[t]
\includegraphics[width=77mm]{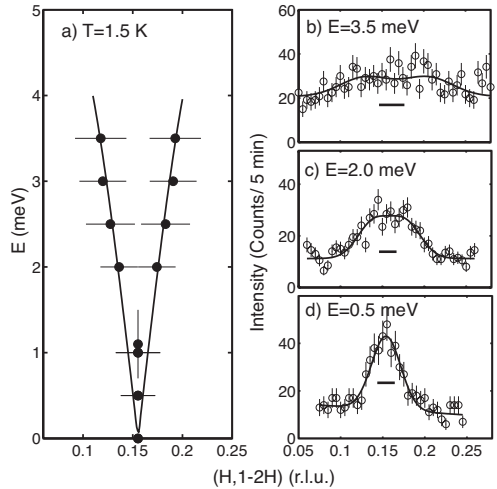}
\caption{The momentum dependence of magnetic neutron scattering measured on SPINS. $a)$ is the dispersion and $b)$ -$d)$ are constant energy scans. Data at $\hbar\omega$=0.5, 1.0, 2.0, and 3.0 meV were taken with E$_{f}$=3.7 meV and  $\hbar\omega$=2.5 and 3.5 meV with E$_{f}$=5.1 meV. The analyzer was configured for 5$^{\circ}$ horizontal acceptance.  The horizontal bars are the resolution full-width.} \label{dispersion}
\end{figure}

	To better understand the origin of the SR order, we examined the spin response to an in-plane magnetic field.  Fig. \ref{field} $b)$ and $c)$ show scans along ${\bf{c^{*}}}$ at 10 T and 0 T.  With increasing field, there is a reduction in the amplitude of the intensity modulation indicating a decrease in the $c$ axis correlations (reflected by $\alpha$ in Eq. 1).  Fig. \ref{field} $a)$ shows the field dependence of the intensity at two points along the L-scan: L=0 where the intensity increases and L=1 where it decreases with field.  Quantitatively, the fit to Eq. \ref{formula1} represented by the solid lines in panels $b)$ and $c)$ give $\alpha(0T)/\alpha(10T)$=1.5(2).  Therefore, inter-plane correlations are suppressed in a magnetic field.  In substitutionally disordered magnets, the application of a magnetic field produces a random field which couples linearly to the order parameter.~\cite{Fishmann79:12}  A possible explanation for the high field data is that in-plane pinning is strengthened in a field at the expense of inter-plane correlations. 

	We now examine magnetic excitations in the low-T limit from the SR ordered frozen state at T=1.5 K.  Figs. \ref{dispersion} (b)-4(d) display constant energy cuts along the (HH1) direction for energies 0.5 meV, 2 meV and 3.5 meV.  The inelastic spectrum is summarized in Fig. \ref{dispersion} in the (HK0) scattering plane.  For increasing energy transfer the correlated magnetic scattering broadens in momentum and then disappears at higher energy transfers.  The data can be described by two peaks displaced symmetrically from ${\bf{Q}}$=(0.155,0.155) and separating in proportion to energy  corresponding to a velocity $\hbar c$=29(3) meV \AA.  Fig. \ref{dispersion} $a)$ shows the peak positions at each energy.  Owing to the short-range correlations observed, the peaks are significantly broader than resolution.

	The spin-wave velocity, $\Theta_{CW}$ and $\delta$ can be described in terms of a weak NN ferromagnetic coupling ($J_{1}$) and a dominant third NN coupling ($J_{3}$).  The peak position ${\bf{Q_{0}}}$ is associated with the minimum in the exchange field defined as $J_{q}=\sum_{{\bf{r_{n}}}}J_{n}e^{(-i{\bf{q}}\cdot{\bf{r_{n}}})}$ where $J_{n}$ is the exchange constant for a neighoring spin located at a distance ${\bf{r}}$.  The spin-wave dispersion can be obtained from~\cite{Jensen:book}  

\begin{eqnarray}
E^{2}(q)={{\langle S \rangle}^{2}}(J_{Q_0}-J_{q}) \left( J_{Q_0}-{1\over 2}{(J_{Q_0+q}-J_{Q_0-q})} \right).
\end{eqnarray}

\noindent $q$ is defined as the displacement in momentum away from the ordering wavevector at $Q_{0}$ and $\langle S \rangle$ sublattice magnetization.  To reproduce both ${\bf{Q_{0}}}$=(0.155(3),0.155(3)), the spin-wave velocity $\hbar c$=29(3) meV \AA, and $\Theta_{CW}$=80(2) we find $J_{3}$=2.8(6) meV and $J_{1}$=-0.35(9) $J_{3}$. These values result in the curve in Fig. \ref{dispersion} $a)$, give $\Theta_{CW}$ = -87 K and ${\bf{Q_{0}}}$=(0.153,0.153), consistent with experiment.   The weakly ferromagnetic $J_{1}$ is compatible with the cuprates for Cu-O-Cu bond angles less than $\sim$ 100$^{\circ}$.\cite{Shimizu03:68,Mizuno98:57}  The NN Ni-S-Ni bond angle in NiGa$_{2}$S$_{4}$ is $\sim$ 97$^{\circ}$.

	Quasi-2D gapless spin waves are also inferred from heat capacity measurements.~\cite{Nakatsuji05:309} In the low $T$ limit $C/R=(\nu n 3^
{3/2}a^{2}k_{B}^{2}/(2\pi c^{2} \hbar^{2}))\xi(3) T^{2}$, where $a$ is the lattice constant, $k_{B}$ is the Boltzmann constant, and $\xi(3) \sim 1.202$. From the low-T C$_{m}$ we obtain $ \hbar c=$5.3 $\sqrt{n\nu}$ meV \AA.  Assuming, in any given volume element, there are only 2 magnetic critical wavevectors in the Brillouin zone ($n$=2) and the peak in Fig. 1 (d) reflects easy plane anisotropy ($\nu$=1) we find $\hbar c \sim$ 8 meV \AA.  The discrepancy between C$_{m}$ and neutron data could result from a range of velocities in the disordered state or a slow mode that does not contribute to neutron scattering.~\cite{Waldt98:2,Podolsky09:79} 

	We have shown that NiGa$_{2}$S$_{4}$ is a 2D triangular lattice antiferromagnetic which does not develop long range order down to $T/\Theta_{CW} \sim 2 \%$ even in high quality single crystals. Instead spin-freezing occurs for T$\sim$ 10 K and the low $T$ state consists of 2D SR incommensurate order. Many of the physical properties including the incommensurability, the spin-wave velocity and the Curie-Weiss temperature can be accounted for by strong third NN antiferromagnetic and weak NN ferromagnetic exchange.  This model, however, contrasts with the lack of long-range spin order. Quantum fluctuations, and/or anomalous domain dynamics \cite{Kawamure07:76,Bhatt06:74} with dilute strong pinning \cite{Stoudenmire09:79} are avenues that should be further explored to determine why long range order does not occur in high quality single crystalline NiGa$_{2}$S$_{4}$.  Irrespective of the eventual resolution of that puzzle, our results indicate that the S=1 antferromagnetic on a triangular lattice is a marginal case and the collective properties cannot be understood in terms of conventional classical theory as in large-S systems.   

	Discussions with L. Balents and M. P. A. Fisher are gratefully acknowledged. Support was provided by NSERC of Canada, a Grant-in-Aids for Research (17071003,18684020,19052003) from JPSJ and MEXT, Japan,  and the NSF through DMR-0306940 and 0706553.

\end{document}